\documentclass[conference,a4paper]{IEEEtran}
\usepackage{xcolor}
\ifCLASSINFOpdf
 \usepackage[pdftex]{graphicx}
\else
\usepackage[dvips]{graphicx}
\fi
\ifCLASSOPTIONcompsoc
 \usepackage[caption=false,font=normalsize,labelfont=sf,textfont=sf]{subfig}
\else
 \usepackage[caption=false,font=footnotesize]{subfig}
\fi
\usepackage{siunitx}
\usepackage{url}
\usepackage{amsthm}
\usepackage{mathtools}
\usepackage{array}
\usepackage{cite}
\usepackage{tabularx}
\usepackage{amsmath,amssymb,amsfonts}
\usepackage{algorithmic}
\usepackage{graphicx}
\usepackage{textcomp}
\usepackage{hyperref}
\usepackage{bm}
\usepackage{siunitx}
\usepackage{dirtytalk}
 \usepackage{multicol}
\usepackage{breqn}
\begin{document}

\title{Antenna's Performance in Microwave Imaging of Stratified Media}
\author{\IEEEauthorblockN{
Adel Omrani\IEEEauthorrefmark{1},  
Sajjad Sadeghi\IEEEauthorrefmark{2}
}                                 
\IEEEauthorblockA{\IEEEauthorrefmark{1} High-Frequency Engineer, Karlsruhe, Germany\\
\IEEEauthorblockA{\IEEEauthorrefmark{2} Institute of Electronics, Graz University of Technology,  Graz, Austria  }
Email: \href{mailto:adelomrani69@gmail.com}{adelomrani@gmail.com}}
}
\maketitle
\begin{abstract}
Numerous types of antennas have been employed for microwave imaging of stratified media for ground penetrating radar (GPR), through-the-wall-radar imaging (TWRI), etc. This letter aims to investigate the impact of the different antennas with their characteristics on the image reconstruction of those media. Hence, three types of antennas, including horn antennas, open waveguide and Vivaldi antennas, are chosen as almost directional antennas, operating at X-band (\textbf{8-12\,GHz}). The antenna's far-field and near-field characteristics are analyzed. A diffraction tomography (DT)-based algorithm is used to reconstruct the target location within the stratified media using monostatic and multistatic data. It is observed that the more directional antennas provide a better-reconstructed image with less shadowing image of the stratified media.
 \end{abstract}

\vskip0.5\baselineskip

\section{Introduction}
Microwave imaging of stratified media has found many applications in different areas, such as GPR, TWRI, and industrial applications, to name a few. Many attempts have been made to develop the proper imaging algorithms, such as diffraction tomography (DT), time-reversal imaging (TRA), and synthetic aperture radar (SAR), to locate the target's location inside the stratified media with the use of different antennas \cite{Adel1, Adel2, Sadeghi_Faraji-Dana_2021, AdelTR}. Antennas are an inevitable part of a microwave tomography (MWT) system to provide a high-quality image of the inspected medium and deliver a compact MWT system. Several omnidirectional and directional antennas have been employed in GPR, TWRI, and industrial applications in the near-field, far-field, or radiative near-field region. Depending on the applications, one can choose from numerous antennas.\newline
For example, in \cite{MUDT_AP, adelMUDT, 9696228, yadav2022neural, MohammadjavadZakeri, Zakeri02092025}, the open-waveguide X band antenna (WR90), positioned in the radiative near-field range, was used to reconstruct the location of the moisture inside a polymer foam \cite{Zhang} while the MWT was working near a high-power microwave drying system operating at 2.45 GHz. The choice was made to suppress the unwanted power leakage from the microwave drying system. Horn antennas (in far-field \cite{Sadeghi_Dort} or near-field \cite{Kai}) are widely used in several scenarios in the literature. These antennas are moderately bulky because of their structure and may not lead to a compact MWT system. However, their performance is satisfactory. Two-dimensional antennas such as Vivaldi and patch are also used in microwave imaging \cite{SadeghiV}.\newline
\indent In \cite{Abosh}, the antennas with omnidirectional and directional antennas are compared for microwave head imaging. In this work, we compare the performance of several directional antennas, including the Horn, open waveguide antenna, and Vivaldi, with different degrees of directivity. The antenna array is positioned at the radiative near-field to increase the signal-to-noise (SNR) ratio of the MWT system. In this regard, both the far-field and near-field characteristics of the mentioned antennas are studied to analyze the performance of the reconstructed image. 
\begin{figure}[!hbt]
\centering
 \includegraphics[width=0.4\textwidth]{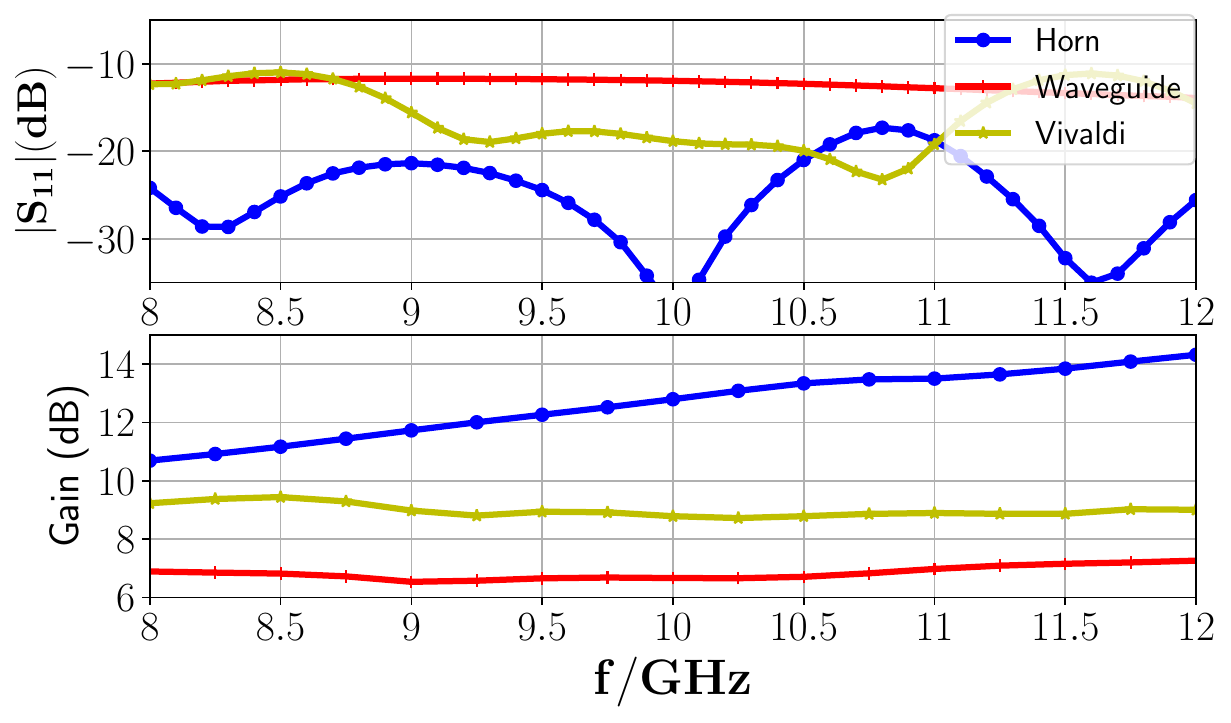}
\caption{Reflection coefficients (top), and Gain (bottom) performance of the horn, waveguide, and Vivaldi antennas in X-band frequency range. }
   \label{S11Gain1}
 \end{figure}
\section{Far-field Characteristics}
\begin{figure}[!hbt]
\centering
\includegraphics[width=0.24\textwidth]{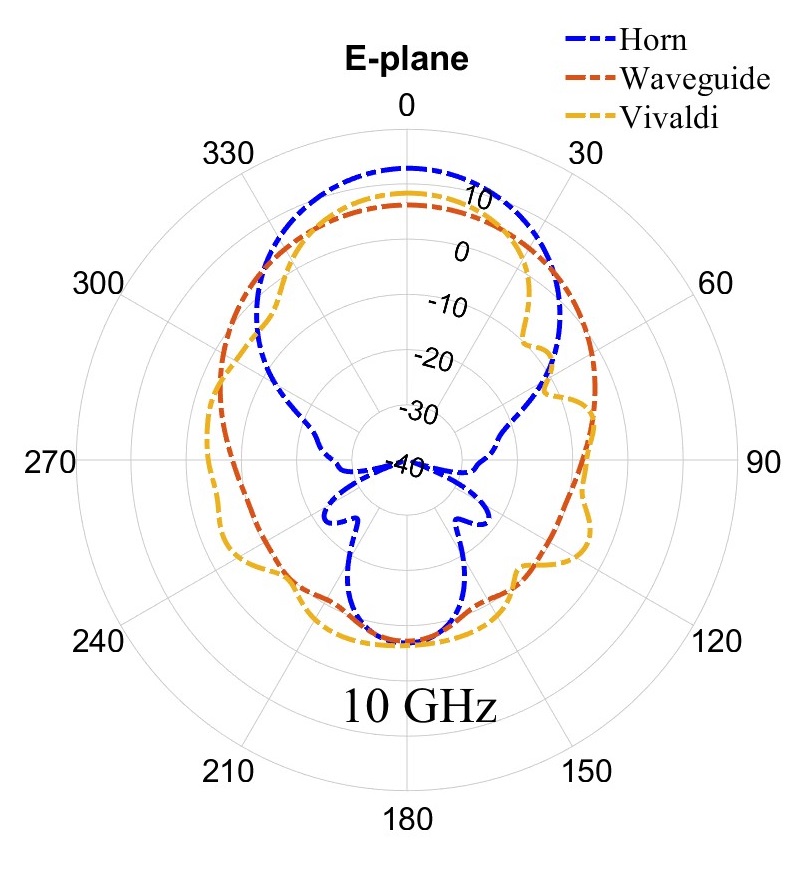}
\includegraphics[width=0.24\textwidth]{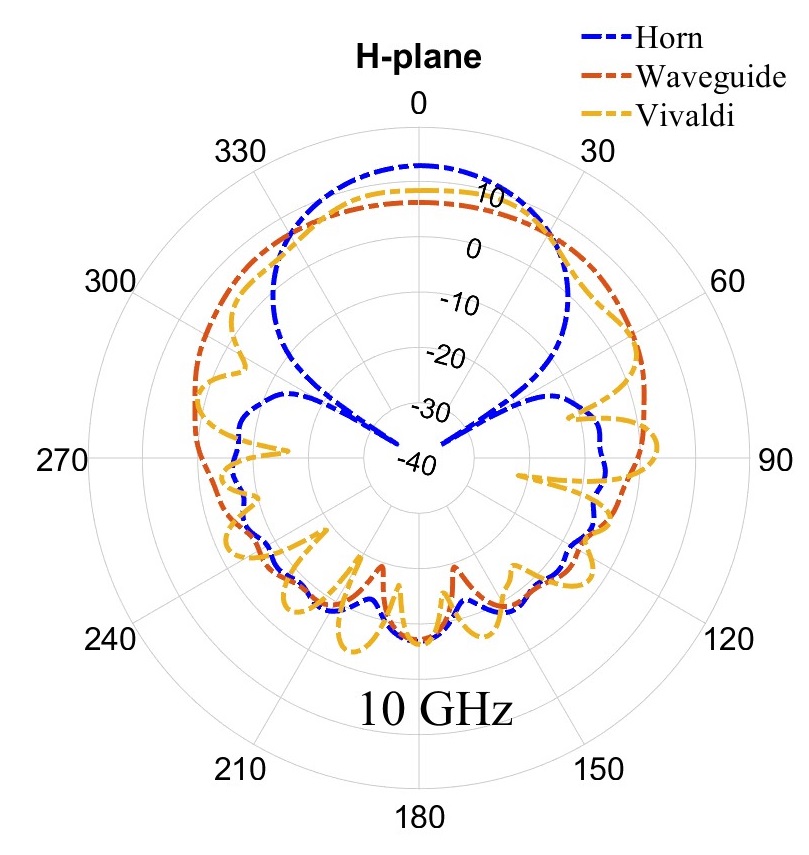}
\includegraphics[width=0.24\textwidth]{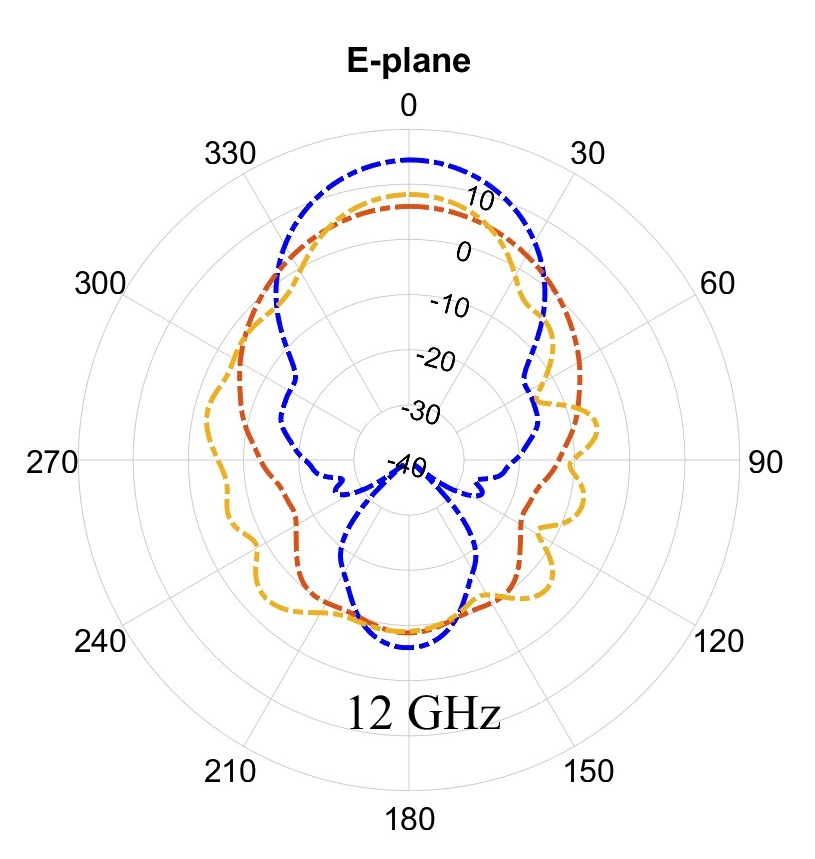}
\includegraphics[width=0.24\textwidth]{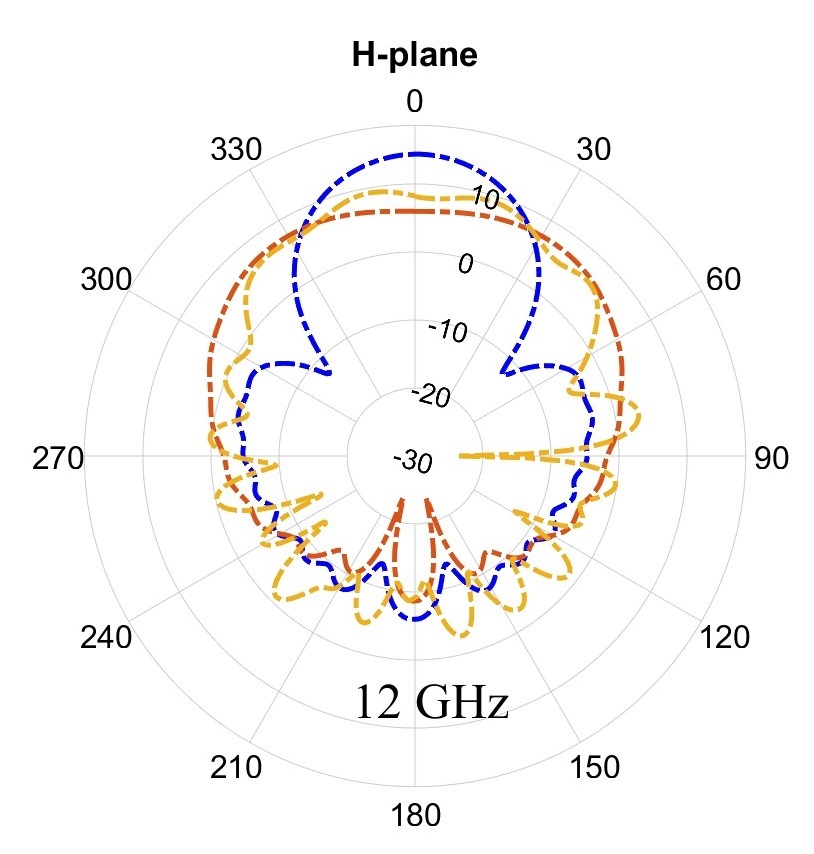}
\caption{E-plane (left), and H-plane (right) radiation pattern of the horn, waveguide, and Vivaldi antenna at \SI{10}{GHz} and \SI{12}{GHz}. }
   \label{Rad1}
 \end{figure}

The reflection coefficients and gains in antenna frequency are presented in Fig.\ref{S11Gain1} (top) and \ref{S11Gain1} (bottom), respectively. The gain of the horn antennas is higher than open waveguide and Vivaldi antennas, Providing more focus electric field strength. The gain of both the Open waveguide and Vivaldi antennas is almost constant and smooth over the frequency band. The radiation pattern of the antennas at the frequencies of \SI{10}{GHz} and \SI{12}{GHz} is plotted in Fig. \ref{Rad1} for the E-plane and the H-plane. Obviously, the radiation pattern of the horn antennas is narrower than that of open waveguide and Vivaldi antennas. Both open waveguide and Vivaldi antennas have almost the same radiation pattern at both planes. In this case, it is expected that a more directional antenna delivers a higher-quality reconstructed image when the mono-static configuration employed for the image reconstruction. More spectral components of less directional components can be recorded by adjacent antennas, which implies that the spectral components are indiscernible. 
 
 \section{Near-field Characteristics}
\begin{figure}[!hbt]
\centering
 \includegraphics[width=0.45\textwidth]{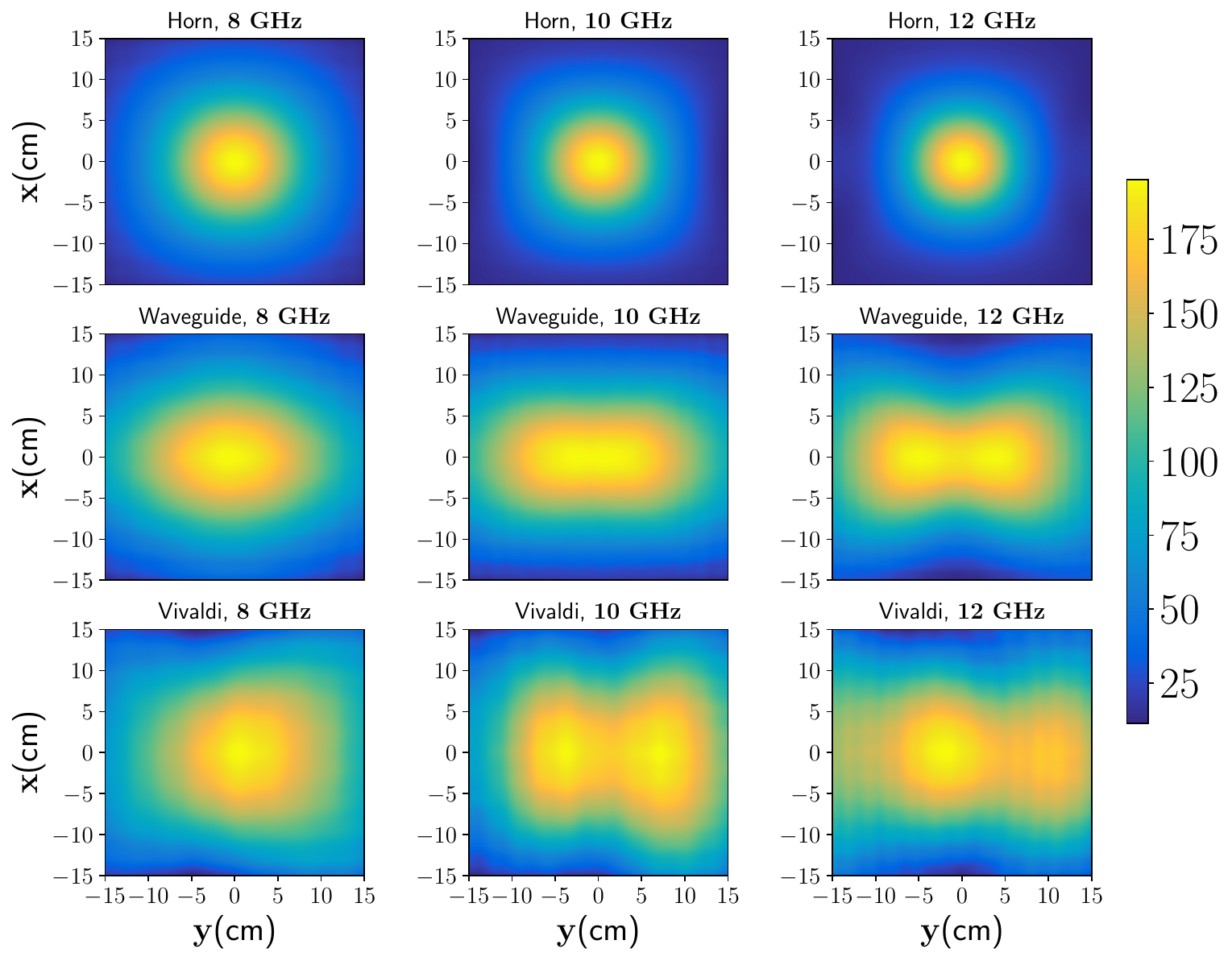}
\caption{The electric field strength of the horn, waveguide, and Vivaldi antennas in the near-field region.}
   \label{FullwaveEfield}
 \end{figure}

\begin{figure}[!hbt]
\centering
 \includegraphics[width=0.45\textwidth]{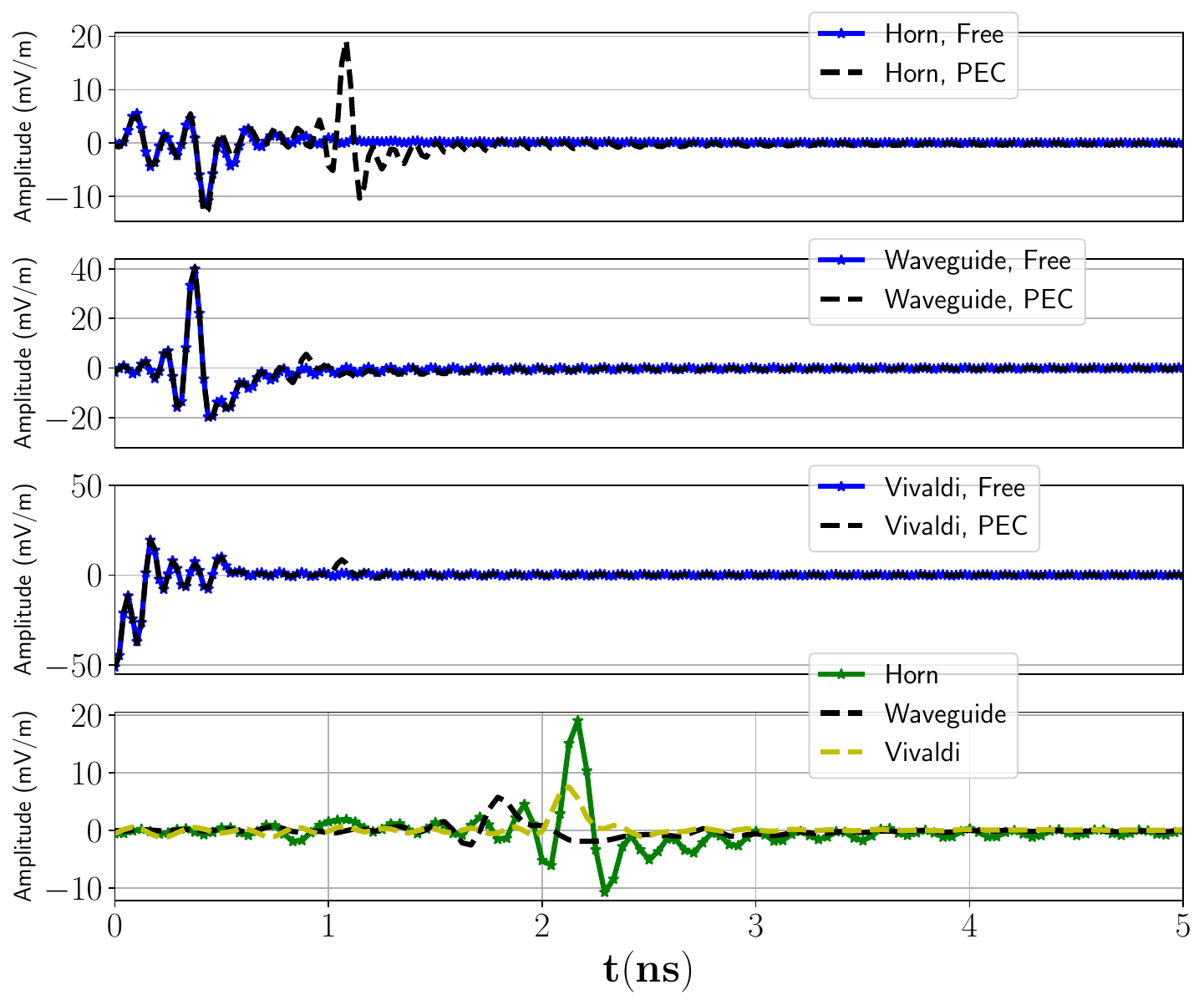}
\caption{The time domain response of of the horn, waveguide, and Vivaldi antennas.}
   \label{TDR}
 \end{figure}

 To better address the effect of the radiation pattern at the near-field region, the electric field strength is depicted at the distance of \SI{12}{cm} from the antennas in $xz$-plane at the different frequencies, as shown in Fig. \ref{FullwaveEfield}. As can be perceived from Fig. \ref{FullwaveEfield}, by increasing the frequency, the open waveguide and Vivaldi antennas start radiating with two unequally strong E-field concentrations along the feeder. This indicates a bi-static imaging algorithm might perform better than a mono-static imaging algorithm since part of the spectral component is received by the adjacent elements. However, for horn antennas, the electric field distribution on the directional antenna is uniform on the front side, resulting in stronger near-field radiation compared to those antennas. It can be concluded that, for a uniform distribution of electric field over the operating frequency band, the mono-static imaging algorithms provide fewer artifacts compared to the nonstable distribution electric field in the same bandwidth, as will be shown later. 
  \newline
\indent Since antennas are intended to operate from close proximity to stratified media, near-field pulse quality analysis of the transmitted pulses is useful to examine the efficacy of the antennas. To that end, individual antennas are vertically placed in CST MWS software in free space and also near the PEC plate at the distance of \SI{20}{cm} from the antennas. The response in the time domain of three antennas is plotted in Fig. \ref{TDR}. To do that, Hermitian processing is applied to the frequency domain data, and then the IFFT is applied. The time domain response of the antennas that perform in free space and near a PEC plate is shown in Fig. \ref{TDR}. As can be seen, in the interval of \SI{0}{ns} to less than \SI{1}{ns}, the response in the time domain of each antenna with and without the PEC plate is the same. This is due to the internal reflections of the antennas \cite{internal}. After extracting the internal reflection of the antennas from the response, the response in the time domain of the antennas near a PEC plate is plotted in Fig. \ref{TDR}(bottom).
As shown in Fig. \ref{TDR} (bottom), intrinsic reflections of the antenna and pulse distortions create additional spectral content. These unwanted components mix with the true scatterer response, leading to increased aliasing effects in the reconstructed image.
 \section{Imaging Performance and Conclusion}
\begin{figure}[!hbt]
\centering
 \includegraphics[width=0.4\textwidth]{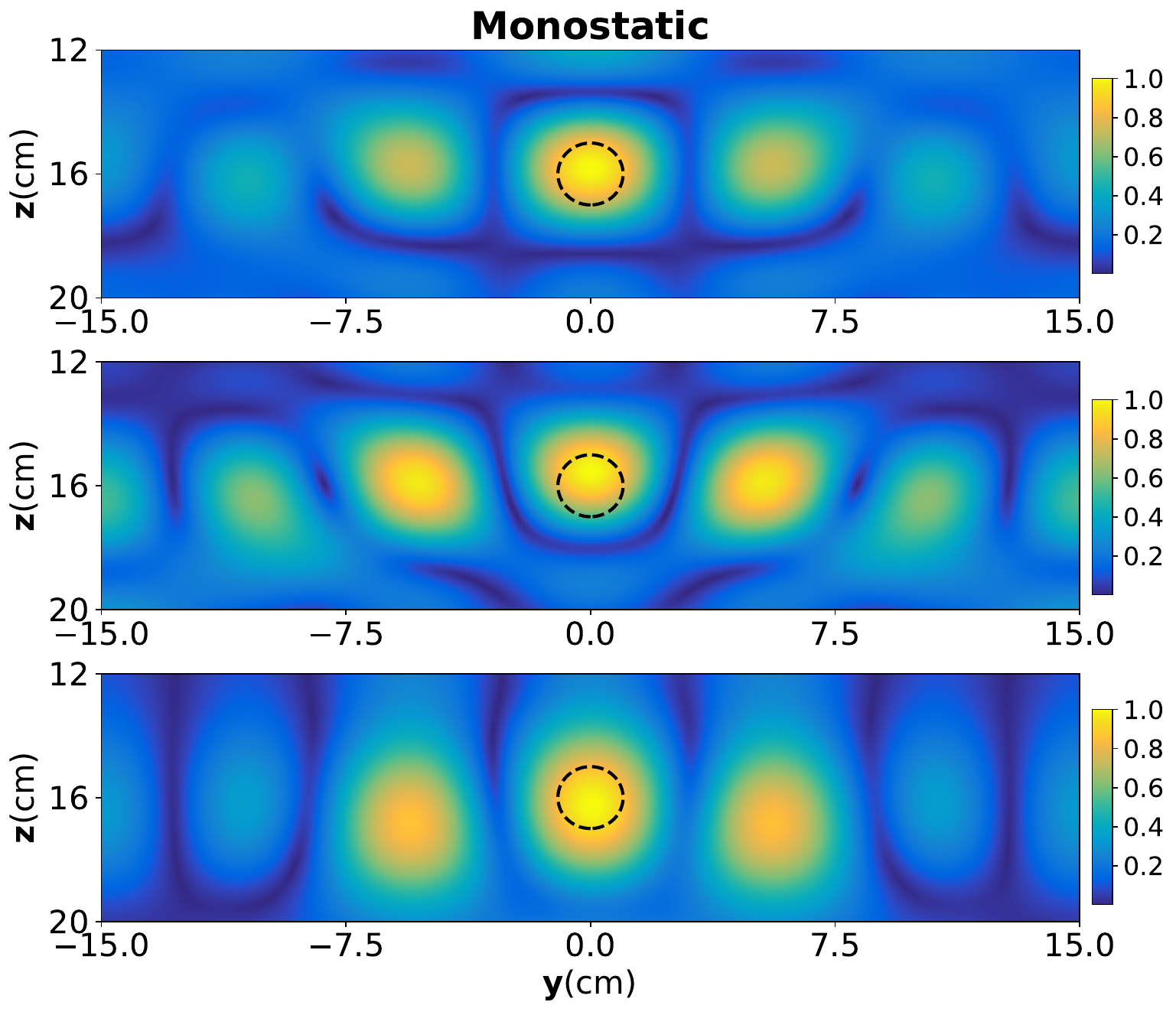}
 \includegraphics[width=0.4\textwidth]{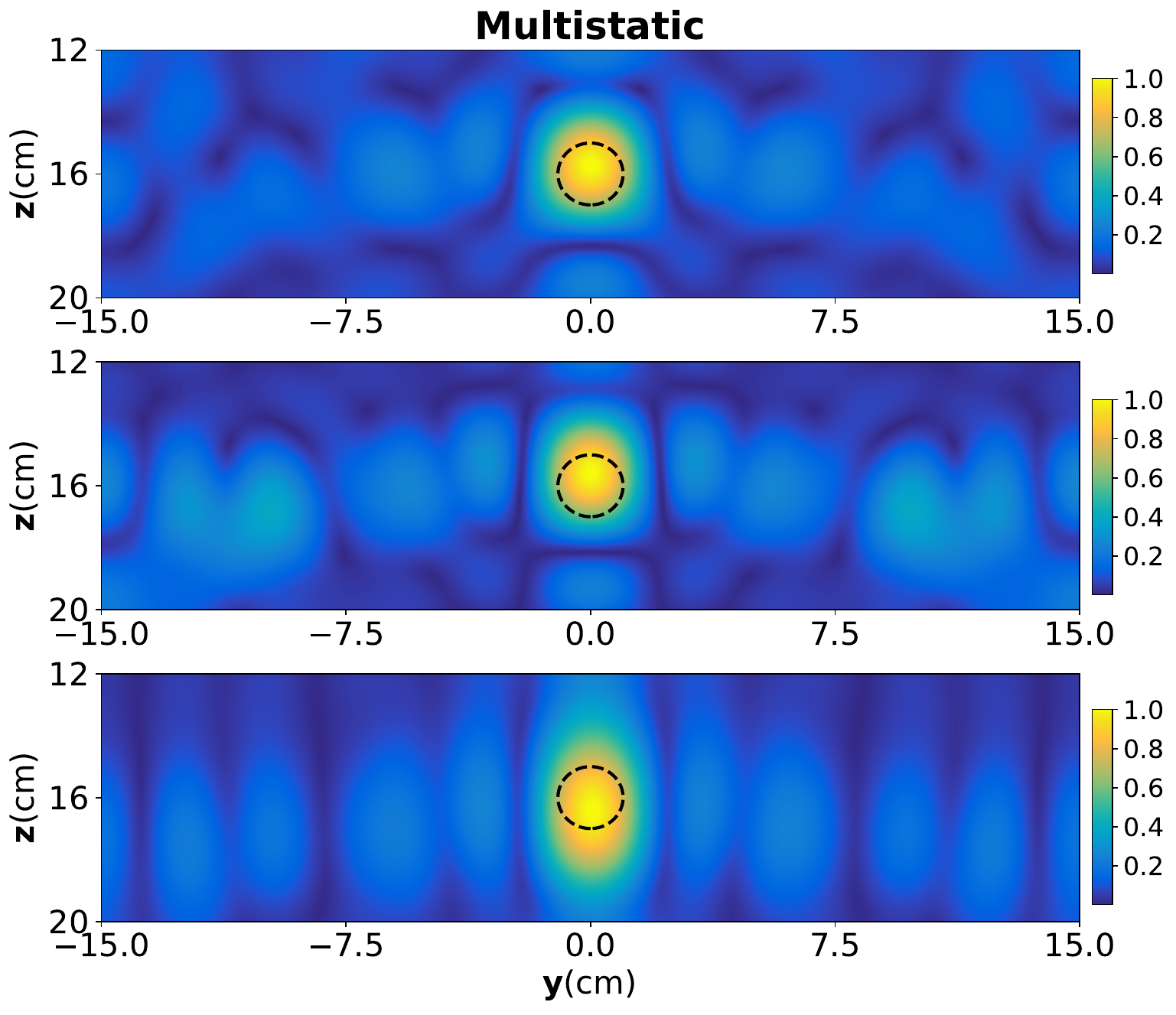}
\caption{Reconstructed image using the UDT(top), and MUDT (bottom) algorithms with horn, waveguide, and Vivaldi antennas. }
   \label{S11Gain}
 \end{figure}

 Fig.  \ref{S11Gain} illustrates that a directive antenna results in less aliasing compared to a broader antenna, but it does not guarantee aliasing-free images. To achieve aliasing-free images, bistatic or multistatic arrangements are required. The use of non-diagonal elements of the scattering matrix significantly enhances the quality of reconstructed images in all scenarios. Additionally, if non-diagonal elements are not feasible, a directive antenna may introduce more artifacts in the reconstructed images compared to omnidirectional antennas. As shown in Fig. \ref{FullwaveEfield}, using higher frequencies for image reconstruction with Vivaldi and waveguide antennas can increase the aliasing effect. This occurs because the spectral components, particularly in bistatic and multistatic algorithms, are influenced by non-uniform field strength, and components originating from one position may also be received by other antennas in the array. In other words, by analyzing the electric field strength, the appropriate frequency range for the imaging algorithm can be selected.

\IEEEtriggercmd{}
\IEEEtriggeratref{4}

\bibliography{references}

\end{document}